\documentclass{article}
\usepackage{spconf,amsmath,graphicx}
\usepackage[utf8]{inputenc} 
\usepackage[T1]{fontenc}    
\usepackage{hyperref}       
\usepackage{url}            
\usepackage{booktabs}       
\usepackage{amsfonts}       
\usepackage{nicefrac}       
\usepackage{microtype}      
\usepackage{xcolor}         
\usepackage{graphicx}
\usepackage{wrapfig}
\usepackage{multirow}

\title{Audio Deepfake Detection System with Neural Stitching for ADD 2022}
\name{Rui Yan, Cheng Wen, Shuran Zhou, Tingwei Guo, Wei Zou, Xiangang Li}
\address{Beike, Beijing, China}
\email{\{yanrui020, zouwei026, lixiangang002\}@ke.com}
\begin{document}
%
\maketitle
\begin{abstract}
This paper describes our best system and methodology for ADD 2022: The First Audio Deep Synthesis Detection Challenge\cite{Yi2022ADD}. The very same system was used for both two rounds of evaluation in Track 3.2 with similar training methodology. The first round of Track 3.2 data is generated from Text-to-Speech(TTS) or voice conversion (VC) algorithms, while the second round of data consists of generated fake audio from other participants in Track 3.1, aming to spoof our systems.
Our systems uses a standard 34-layer ResNet \cite{szegedy2017inception}, with multi-head attention pooling \cite{india2019self} to learn the discriminative embedding for fake audio and spoof detection. We further utilize neural stitching to boost the model's generalization capability in order to perform equally well in different tasks, and more details will be explained in the following sessions. The experiments show that our proposed method outperforms all other systems with 10.1\% equal error rate(EER) in Track 3.2.
\end{abstract}
\section{Introduction}
\label{sec:intro}
Deep learning technology has empowered many speech applications and made them widely accessible over the last few years. The models can generate realistic speech that most people will find it nearly impossible to distinguish it from the real ones. Meanwhile, users can easily have a customized TTS voice with just few simple utterances as input and a mobile phone. This has brought a great challenge to anti-spoofing researchers. Previous works\cite{khanjani2021deep,todisco2019asvspoof,chen2020generalization} have shown that automatic speaker verification systems (ASV) plays a key role in many of the anti-spoofing scenarios. However, in realistic situations, a very complicated background noise, reverberations or a mixture of real-fake clips might leads to a total failure of the system. Also, the recent leap of VC and TTS technologies \cite{kong2020hifi,ren2020fastspeech,zhao2021towards,ding2021adversarial} brought the authenticity of generated speech to another level, which makes anti-spoofing nowadays even harder.

Audio Deep Sythesis Detection 2022 allows researchers to improve their anti-spoofing system by providing more realistic data, more diversified and challenging scenarios. In low-quality fake audio detection (LF) track, data is mixed with real and spoof-attack generated (VC and TTS) data. In audio fake game detection task (FG-D), the evaluation set is composed of the data from the very same generation task generated from other competitors, which creates an adversarial game among participants.These two tasks represent a more realistic situation of what anti-spoofing researchers need to deal with day to day.

To address these problems, this paper proposes a system and training methodology to detect the spoofing anomalies more effectively. We also analyse how different type of data augmentation through speed, volume, noise, spectrum and codec methods can greatly increase the overall performance of the whole system. The following sections of this paper is organized as follows. Section 2 introduces the model structure and how neural stitching helps the final result. Section 3 will go through the features, data augmentation and other details of the experiments. And section 4 concludes the paper.

\section{System Description}
\label{sec:format}
In this section we will firstly present the framework of the system we submitted, and then describe the reason that it was chosen for the fake speech detection task.
\subsection{Established Models}
Residual network(ResNet)\cite{szegedy2017inception} is commonly used in deep learning tasks, it captures more discriminative local descriptors and then aggregate them to generate final fixed dimensional utterance-level embeddings. The embeddings are then fed into a classifier which tells whether the input audio is a spoof attack or genuine. 

The proposed model is composed of two networks: an embedding network and a classification network. The embedding network is essentially a ResNet-34 and attention pooling is then applied to aggregate the frame-level output. The output's first and second order moments are calculated and pooled together to obtain the final utterance-level representation. The pooled representation are then fed into the classification model, which consists of two fully-connected layer and a 2-dimensional softmax layer, where each node represents the bona fide and fake audio class, respectively.

The activation function is relu in embedding network, where mish\cite{misra2019mish} is adopted in classification network.
Benefiting from it`s properties like being unbounded above, bounded below, smooth and non-monotonic, mish has achieved generally better performance than relu in various classification tasks.
Focal loss\cite{lin2017focal} was used for training. The solid line in Fig.1 shows the data flow for training.
\begin{figure}[!h]
  \centering
  \includegraphics[trim = 0mm 0mm 0mm 0mm, width=0.45\textwidth]{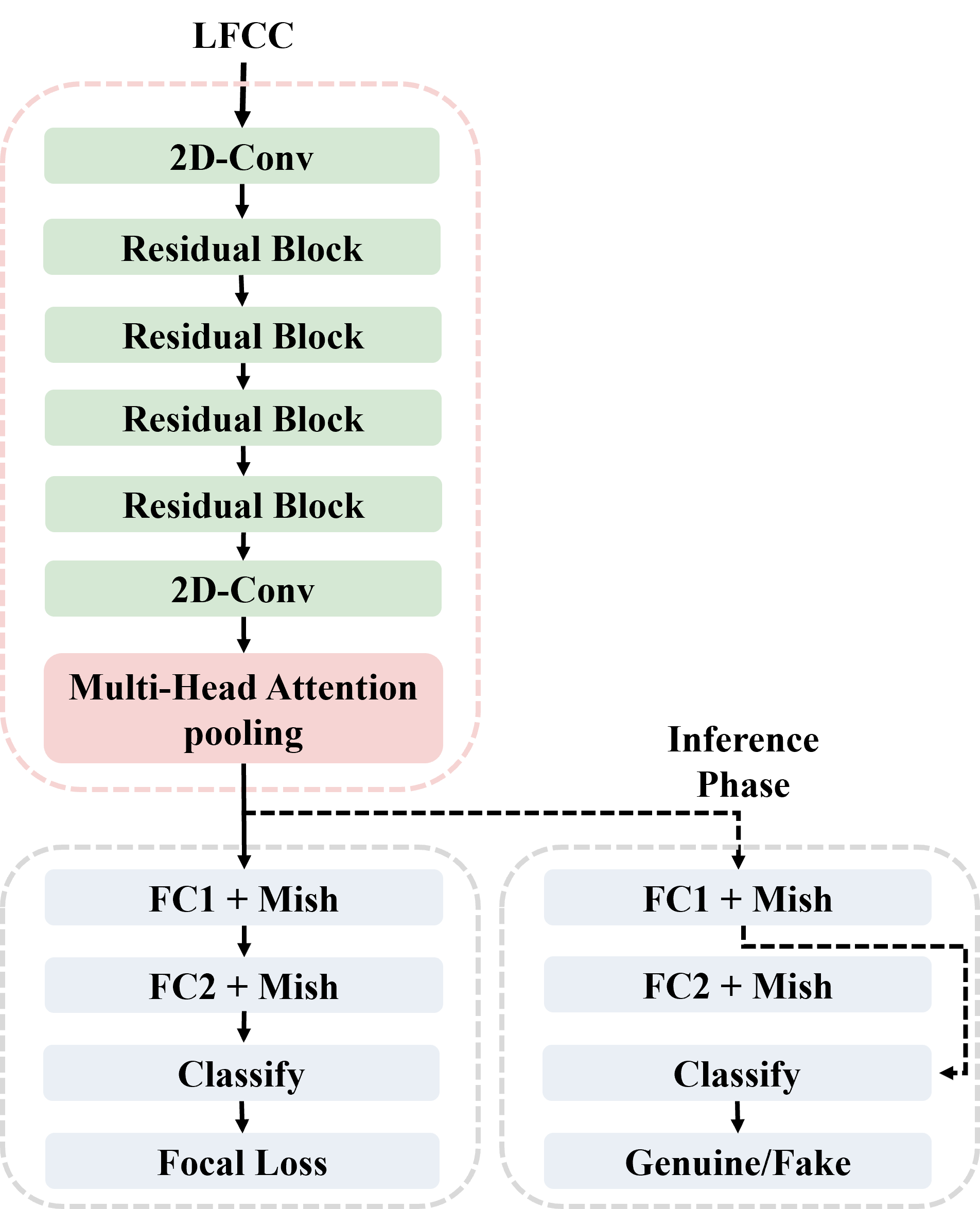}
  \caption{General architecture of the ResNet-based neural spoof detection system.(The solid line is the data flow for training, the dotted line is the data flow for inference.)}
  \label{fig:dsr}
\end{figure}
\subsection{Neural Stitching}

In a previous study of features in the area of computer vision\cite{lenc2015understanding}, researchers studied three key mathematical properties of different representations: equivariance, invariance and equivalence. To study CNN representation's equivalency, two different CNN models trained on the same dataset are decomposed into four parts, $\phi = \phi_1 \circ \phi_2$ and $\phi^{\prime} = \phi_1^{\prime} \circ \phi^{\prime}_2$, and the goal is to find a mapping of $E_{\phi_1->\phi_1^{\prime}}$ that allows $\phi^{\prime}_2 \circ E_{\phi_1->\phi_1^{\prime}} \circ \phi_1$ to perform the same as the original $\phi_2^{\prime} \circ \phi^{\prime}_1$. The $E_{\phi_1->\phi_1^{\prime}}$ mapping is here interpreted as the stitching layer, and it can be implemented as a bank of filters or transformation. After the stitching operation, image classification performance is dramatically better than before, indicating that some extent of the feature exchange among different networks trained on the same dataset could be helpful. 

Inspired by this idea, we would like to see how the stitching works in a single model by decomposing the model layers into different stitching parts. Some of the previous work indicate that shallow layers of network tends to learn more general representation and deep layers of network focus more on specific tasks\cite{zeiler2014visualizing}. Thus we cut off one deep layer during the inference stage and stitched the classification layer directly to the output of ResNet, and the result was surprisingly good. The newly stitched structure increases the model robustness, while at the same time could reduce overfitting. The result in Table 4 shows a very promising improvement in EER after we stitched the output of first dense layer directly to the output layer. The dotted line in Fig.1 shows the inference flow.

\section{Experimental And Result}

\subsection{Datasets and Feature}
In this experiment, we combine the training and verification set to obtain a total of 55K utterances as training set. Each utterance is then divided into fixed length segments with 50\% overlap. During the training, 4000 segments are randomly selected as the verification set and the rest as the test set. In the subsequent experiment, we take the Track 1 adaptation set and Track 3.2 adaptation set as the evaluation set.

\subsubsection{Feature Selection}
In our experiment, we test three types of acoustic features which are proved to have shown good performances in previous fake speech detection research.

\begin{itemize}
  \item Linear frequency cepstral coeffificient (LFCC)\cite{sahidullah2015comparison}: the LFCC is extracted from the magnitude spectra, which is followed by a linear scale filterbank.
  \item DCT-DFT spec: the DCT-DFT spec is extracted similarly to LFCC, but instead of applying linear scale filterbank, the DCT-DFT spec directly uses the log-compressed magnitude spectra. 
  \item Log-linear filterbank energy (LLFB)\cite{sahidullah2015comparison}: the LLFB is extracted in the same pipeline to the LFCC, but no discrete cosine transformation (DCT) operation is applied. 
\end{itemize}

According to ASVspoof 2019 \cite{todisco2019asvspoof} results, static features are better than dynamic features in fake speech detection task, so we only tested static features in this experiment. At the same time, in order to verify the feature robustness, we train with clean data and test in two test sets. One is the noisier Track 1 adaptation set, and the other is the cleaner Track 3.2 adaptation set. The results are shown in the Table 1.

\begin{table}[!h]
  \caption{Comparison of EER (\%) performance on different feature.(Ta1 is a Track 1 adaptation set, Ta3 is a Track 3.2 adaptation set)}
  \centering
  \resizebox{0.5\textwidth}{!}{
  \begin{tabular}{llc}
   \begin{tabular}{cccccc}
  \toprule
   \multicolumn{1}{c}{feature}
    & \multicolumn{1}{c}{dim}
    & \multicolumn{1}{c}{nfft}
    & \multicolumn{1}{c}{Ta1 EER(\%)}
     & \multicolumn{1}{c}{Ta3 EER(\%)}\\
    \midrule
    DCT-DFT spec & - & 512 & 25.57 & 8.73\\
    \cline{2-5}
    \multirow{4}{*}{LFCC} & 80  & 512 & 21.29 & 1.01\\
    & 80 & 1024 & \textbf{18.43} & 0.67\\
    & 60 & 1024 & 19.71 & 0.61\\
& 40 & 1024 & 20.71 & 0.82 \\
    \cline{2-5}
    LLFB & 80  & 1024 & 20.57 & \textbf{0.56}\\
\bottomrule
\label{table:cer}
 \end{tabular}
\end{tabular}
}
\end{table}

As can be seen from Table 1, when the dimension of static LFCC feature is 80 and the FFT bins is 1024, the model achieves the best result in Track 1 adaptation set. LFCC feature and LLFB feature have little difference in performance on Track 3 adaptation set. Considering the robustness of the features, we used LFCC features uniformly in the following experiments.

\subsubsection{Data Augmentation}
One of the main objectives of ADD challenge is to develop counter measurements that are robust to noises, channel effects, codecs and out-of-domain data. Data augmentation plays an essential role in achieving that goal, especially when those effects or data distribution are not present in the training data. This part describes our approaches to data augmentation for Track 3.2 tasks.

Due to that the Track 3.2 test audio was relatively clean, we only add a single disturbance to each audio. Two different types of data augmentation are applied to the train dataset.

In the first part, we add distortion and noises to the clean samples, which is similar to the augmentation used in\cite{snyder2018spoken}. The mixing data came from publicly available room impulse responses (RIR)\cite{shinn2005localizing} and MUSAN\cite{snyder2015musan}, which includes noise, music and babble noises. To add more spice, we've also included self-collected background noise for more noise coverage. The noises are mixed with a random Signal-to-Noise ratio (SNR) between 0 dB to 20 dB. In addition, volume also effects the detection performance so we randomly change the volume gain from -10dB to 20dB. At last, a total of five times the size of the original clean data is generated and we random sample 60k out of them.

In the second part, we simulate audio compression effects. All clean audio samples are passed through audio compression algorithms to generate simulations. The compression algorithms include MP3, OGG, AAC and OPUS.  At last, we also do frequency conversion to mock the telephony transmission loss, audio samples are first downsampled to 8kHz and then upsampled back to 16kHz. Another five times of the data is generated and we random select 40k from them.

Table 2 shows a detailed summary of the data augmentation setups and Table 3 shows improvement from data augmentation.

\begin{table}[!h]
  \caption{Summary of Data Augmentation.}
  \centering
  \resizebox{0.5\textwidth}{!}{
  \begin{tabular}{llc}
   \begin{tabular}{cccccc}
  \toprule
   \multicolumn{1}{c}{Approach}
    & \multicolumn{1}{c}{Methods}
    & \multicolumn{1}{c}{Description}
    &\multicolumn{1}{c}{Size}\\
    \midrule
    \multirow{5}{*}{Distortion} & noise & \multirow{3}{*}{MUSAN and self-collected} & \multirow{5}{*}{60k}\\
    & music \\
    & babble \\ 
    \cline{2-3}
    & reverb & RIR \\
    \cline{2-3}
    & volume & -10dB to 20dB \\
    \midrule
    \multirow{5}{*}{Compression} & MP3 & \multirow{4}{*}{Random compression rate} & \multirow{5}{*}{40k}\\
    & OGG \\
    & AAC \\
    & OPUS \\
    \cline{2-3}
    & sample rate & 16kHz -> 8kHz\\
\bottomrule
\label{table:cer}
 \end{tabular}
\end{tabular}
}
\end{table}

\begin{table}[!h]
\vspace{-0.1em}
  \caption{Comparison of EER (\%) performance from data augmentation.((Ta1 is a Track 1 adaptation set, Ta3 is a Track 3.2 adaptation set))}
  \centering
  \resizebox{0.35\textwidth}{!}{
  \begin{tabular}{llc}
   \begin{tabular}{cccccc}
  \toprule
  \multicolumn{1}{c}{}
    & \multicolumn{1}{c}{before (\%)}
    & \multicolumn{1}{c}{after (\%)}\\
    \midrule
      Ta1 EER & 18.43  & \textbf{3.71} \\
      Ta3 EER & 0.67   & \textbf{0.34} \\
\bottomrule
\label{table:cer}
 \end{tabular}
\end{tabular}
}
\vspace{-0.1em}
\end{table}

As shown in Table 3, after adding noise to the training data, EER of track 1 test set decreases dramatically, and that of Track 3 test set also decreases to a certain extent. We use this set for further training.

\subsection{Training Setup}


\subsubsection{Model Structure Comparison}
In previous works\cite{india2019self,snyder2018spoken,cai2018novel,okabe2018attentive,wang2020multi}, a carefully chosen pooling layer shows great impact to the final results. We study and compare five types of pooling layers in our experiments, among which learnable dictionary encoding(LDE)\cite{cai2018novel} pooling and self multi-head attention(MH)\cite{india2019self} pooling show much more potential comparing to others.

Some of the predicted scores are close to either 1 or 0 before neural stitching, which might be caused by overfitting. After we utilize neural stitching here, the model shows much more reasonable results. In the test set, EER improves significantly after neural stitching. Later on, we also perform a series of fine-tuning to achieve our final result. The comparison of pooling layers and stitching is shown in Table 4. It can be seen that the EER on Track 3.2 round 1 test set is significantly reduced after stitching.

\begin{table}[!h]
  \caption{Comparison of EER (\%) performance comparison of pooling layers and stitching. (ST:Statistic pooling\cite{snyder2018spoken}, AT:Attentive statistics pooling\cite{okabe2018attentive}, MRH:Multi-resolution multi-head attention pooling\cite{wang2020multi}, Ta1:Track 1 adaptation set, Ta3:Track 3.2 adaptation set, Test:Track 3.2 round 1 test set. N:normal inference, S:neural stitching inference)}
  \centering
  \resizebox{0.5\textwidth}{!}{
  \begin{tabular}{llc}
   \begin{tabular}{cccccccccc}
  \toprule
   Pooling
    & \multicolumn{2}{c}{Eval(N))}
    & \multicolumn{2}{c}{Eval(S)}
    & Test(N)
    & Test(S)\\
    \cmidrule(r){2-3} \cmidrule(r){4-5} 
     & Ta1 & Ta3 & Ta1 & Ta3\\
    \midrule
    ST           &  4.57   &  0.34   &  8.86   & 0.11  & - & -\\
    LDE                     &  5.42   & 1.01   &   4.42   &  0.0   & 22.74 &  \textbf{13.77} \\
    AT         &  4      & 0.34   &  19.72  & 0.89  & - & -\\
    MH              &   3.71   & 0.34   &  8.43   & 0.11  & 21.14 & \textbf{12.96} \\
    MRH   &  4      & 0.34   &  38.3   & 14.56 & - & -\\
\bottomrule
\label{table:cer}
 \end{tabular}
\end{tabular}
}
\end{table}

\subsubsection{Model Fine-tuning}

In the fine-tuning process, we introduce specAugment\cite{park2019specaugment,zhong2020random} and chunk size as variables to enhance the model robustness. SpecAugment randomly blocks certain percentage of frequency bins or time frames and replaces them with a constant value, acting as a changing T-F mask. It shows a reliable and steady improvement of performance in many of the speech tasks. Fine-tuning results are presented in Table \ref{tab:asr}.



From Table \ref{tab:asr} we can see that chunk size significantly influences the results. It should be set within a reasonable range, which we think may be related to the audio length of the test set. While a large chunk size such as 1500 ms is used, model converges in few epochs and training accuracy increases to 100\%, which shows an obvious overfitting. In addition, although specAugment indeed can improve the robustness of the model, the T/F replacing percentage should not be set too large in the training process if small chunk sizes are chosen. 

\begin{table}[!h]
  \caption{Comparison of EER (\%) with different parameters on Track 3.2 Test.(F: At most F\% of the frequency bins masked with a constant value, T: At most T\% of the time bins masked with a constant value, Rows: The maximum number of frequency mask, Cols: The maximum number of time mask. Chunk: The length of speech sent to the neural network at one time, in milliseconds.)}
  \label{tab:asr}
  \centering
  \resizebox{0.45\textwidth}{!}{
  \begin{tabular}{cccccccc}
    \toprule
    \multicolumn{4}{c}{SpecAugment}  & Chunk & EER(\%) \\
    \cmidrule(r){1-4} 
       F(\%) &  T(\%)  & Rows & Cols\\
    \midrule
     20  & 20 & 2 & 2 & 800  &  12.96\\
	 10  & 10 & 2 & 2 & 800  &  10.64\\
	 10  & 10 & 1 & 1 & 1500 &  44.31\\
     10  & 10 & 1 & 1 & 800  &  10.42\\
     10  & 10 & 1 & 1 & 600  &  \textbf{8.83} \\
     10  & 10 & 1 & 1 & 500  &  17.31\\
    \bottomrule
  \end{tabular}
  }
\end{table}

A model with proper chunk size such as 600 ms and specAugment leads to the best result in 8.83\% in EER.

Finally, on the basis of the optimal model, we modify the overlap ratio of training data from 50\% to 70\%. Then we turn down the learning rate by a factor of 100 and continued training to get the final model.

\begin{table}[!h]
  \caption{Track 3.2 Result}
  \centering
  \resizebox{0.4\textwidth}{!}{
  \begin{tabular}{llc}
   \begin{tabular}{cccccc}
  \toprule
  \multicolumn{1}{c}{}
    & \multicolumn{1}{c}{Round1}
    & \multicolumn{1}{c}{Round2}
    & \multicolumn{1}{c}{Final}\\
    \midrule
      EER(\%) & 8.6 & 11.1 & \textbf{10.1} \\
\bottomrule
\label{table:cer}
 \end{tabular}
\end{tabular}
}
\end{table}
\subsection{Official Result}
In the official evaluation stage, the EER for the first round and second round are 8.6\% and 11.1\% respectively. The final ranking is based on the weighted sum of 40\% of the first round and 60\% of the second round's EER.
The overall ranking is the 1st place in Track 3.2. The results are shown in Table 6.

\section{CONCLUSION}
\label{sec:typestyle}
This paper reports the details of systems we developed during our participation to the ADD Challenge 2022 focusing on detection of speech deepfake attacks. We adopt data augmentation as a primary strategy to improve the robustness of our systems based on known-unknown information as per the challenge evaluation plan. In addition, we propose an innovative network stitching method, which improves the robustness of the model on different distributed test sets, and it shows good performance on all test sets of Track 3.2. Finally, our system outperforms all other proposed systems and won competition with an EER of 10.1\%. In the future, we will test the effectiveness of this method in more speech scenarios, such as speaker recognition and dialect recognition.


\bibliographystyle{IEEEbib}
\bibliography{strings,refs}

\end{document}